# Case Studies of Using the Partial-Structure R1 and the Single-Atom R1 to Assemble Small-Molecule Crystal Structures

Authors

**Xiaodong Zhang[a]\***

[a]Chemistry Department, Tulane University, 6400 Freret Street, New Orleans, Louisiana, 70118, United States

Correspondence email: xzhang2@tulane.edu

**Synopsis**　Pre-knowledge of a crystal structure, including the constituent fragments, the atomic connectivity, and the approximate bond lengths, etc., have been utilized in the partial-structure R1 and the single-atom R1 calculations for assembling small molecule crystal structures.

**Abstract**　This paper demonstrates how pre-knowledge of a crystal structure, including the constituent fragments, the atomic connectivity, and the approximate bond lengths, etc., can be utilized in the partial-structure R1 (pR1) and the single-atom R1 (sR1) calculations. It has been shown that taking advantage of pre-known information the calculations can proceed in an orderly and well-planned manner. Furthermore, in the case of the sR1 calculation, computer time can also be greatly reduced. Because the pR1 calculation is more time-consuming than the sR1 calculation, when there is a choice between the pR1 and the sR1, the former should be avoided. So, the usual strategy of assembling a light-atom-only structure should start with the normal sR1 method to determine a basic framework of the structure, and then uses the connectivity-guided sR1 method to complete the model. However, when the data resolution is low, the first step is necessary to use the pR1 method to assemble the known fragments. For a heavy-atom-containing structure, the correct strategy starts with the normal sR1 method to determine the heavy-atom substructure, and then uses the connectivity-guided sR1 to complete the model.

## 1. Introduction

In many situations, based on the knowledge of the sample preparation, combined with information obtained from other technologies (e.g. NMR, ESI-MS, etc.), a crystallographer has a well-established expectation of what the unknown structure should look like, including the constituent fragments, the atomic connectivity, and the approximate bond lengths, etc. It is well known that such pre-knowledge can be utilized for solving crystal structures. In fact, this is the fundamental basis of the well-established molecular replacement (MR) method (Rossmann & Blow, 1962; Rossmann *et al.*, 1964; Crowther, 1972; Rossmann, 1972, 1990; Bricogne, 1992; Read, 2001; McCoy, 2004; Read & McCoy, 2016; McCoy *et al.*, 2017; Caliandro *et al.,* 2009; Burla, *et al.*, 2020).

A recent paper (Zhang & Donahue, 2024) has introduced the single-atom R1 (sR1) concept, as well as the more general partial structure R1 (pR1) concept. The sR1 and the pR1 are two new model-searching techniques for searching single atoms and fragments, respectively. So far, they have been used for directly solving small-molecule crystal structures (Zhang & Donahue, 2024). In comparison, the traditional model-searching techniques, including the rotation function (Rossmann & Blow, 1962), the translation function (Rossmann *et al.*, 1964), and the log-likelihood-gain on intensities (LLGI, McCoy *et al.*, 2007; Read & McCoy, 2016), etc., usually serve the initial phasing purpose in the MR calculations, and the MR is typically applied in the macromolecular crystallography, though recently, Gorelik *et al.* (2023) have evaluated the MR as implemented in *phaser* (McCoy *et al*., 2007) for small-molecule crystal structure determination from X-ray and electron diffraction data with reduced resolution.

The normal mode of application of the sR1 method does not require other pre-knowledge except for the atomic content of the unit cell, and the calculation proceeds in a blindly and care-free manner (Zhang & Donahue, 2024). This paper demonstrates a new mode of application of the pR1 and the sR1 calculation, in which, the calculation is guided by various pre-knowledge. It has been shown that such guided mode of calculation can proceed in an orderly and well-planned manner.

The same implementation (Zhang & Donahue, 2024) of the sR1 and the pR1 methods is used in this report, except for one improvement: instead of directly feeding the raw reflection intensity $F_o^2(hkl)$ to the calculation, the sharpened intensities are fed (this is a well-known trick in crystallography). A sharpened intensity is calculated as a product of multiplying the raw intensity $F_o^2(hkl)$ with $\exp(2Bs^2)$, where $s=\sin\theta/\lambda$. The parameter B is determined by the Wilson method (see our implementation in the supporting information). We have found that if the sharpened intensities are used, there are fewer ghost atoms during an sR1 calculation. This is understandable, because the sR1 (as well as the pR1) method employs an atomic model with no thermal effect (i.e., using the isotropic displacement parameter $U=0$ Å$^2$), and such a model matches the sharpened data better (because sharpening removes the average thermal effect).

While the sR1 method can do single-atom search, the pR1 method can determine the orientation(s) of a known fragment (Zhang & Donahue, 2024); furthermore, the pR1 method can also position a fragment of a known orientation. More technical details of the implementation of the pR1 method can be found in the supporting information.

The rest of this paper presents three case studies of the pre-knowledge guided pR1 and sR1 calculations. The crystallographic information of the samples used in these studies is shown in Table 1. The paper ends with a discussion of a comparison between the pR1 method and the sR1 method and between the normal and the connectivity-guided sR1 methods.

**Table 1**  Crystallographic information of the samples studied in this paper

| Sample | Formula (excluding H) | Z | non-hydrogen atoms in cell | a(Å) | b(Å) | c(Å) | α(°) | β(°) | γ(°) | space group |
|---|---|---|---|---|---|---|---|---|---|---|
| 1 | $C_{78}$ | 2 | 156 | 12.34 | 15.98 | 16.57 | 114.10 | 90.70 | 103.20 | P-1 |
| 2 | $C_{46}$ | 1 | 46 | 5.95 | 10.80 | 12.97 | 103.77 | 99.95 | 90.46 | P-1 |
| 3 | $IMo_3S_{13}N_3C_{27}$ | 4 | 188 | 11.87 | 37.77 | 13.14 | 90.00 | 92.34 | 90.00 | Pc |

**2. Assembling the structure of sample 1 by the pre-knowledge guided pR1 and sR1 calculation**

The expected structure of sample 1 is shown in step 3 of figure 1. This structure has two molecules in its unit cell. Each molecule consists of 13 benzene rings. The molecule can be viewed as consisting of two identical parts, in which each part is made of a central benzene ring with a side benzene ring attached to each of its six corners. The two parts are connected to each other by sharing one common side benzene ring. Though the whole crystal structure consists of 26 benzene rings, benzene ring is not the largest known fragment in this structure. The largest known fragment is a benzene-star, which consists of a central benzene ring with a C atom sticking out at each of its six corners. The C-C bonds in a benzene-star have a bond length of 1.39 Å. There are four benzene-star fragments. The pR1 method can be used to orient and position these fragments. The four benzene-star fragments share two orientations (each orientation is shared by two fragments). These two orientations (labeled as 0 and 1) are determined by using the pR1 of a free-standing fragment (Zhang & Donahue, 2024), here free-standing means a model consisting of a single fragment with no other known atoms. The first fragment has orientation 0 and is positioned such that its center is at (0.3, 0.3, 0.3). The second and the third fragments share orientation 1, and each is positioned by globally minimizing its pR1 while adjusting $(x, y, z)$ of its center. The fourth fragment has orientation 0 and is positioned by the same method. Step 1 of figure 1 shows the result after orienting and positioning the four benzene-stars. Upon reviewing this result and judging the possible contacting points between neighboring benzene-stars, one concludes that C12 and C31 belong to one side of a benzene ring that bridges two adjacent benzene-stars, and similarly that C22 and C45 belong to one side of another benzene ring that bridges the other two benzene-stars. However, C12 and C31 are too far away and are not bonding to each other, and the same with C22 and C45. For this reason, C12, C31, C22 and C45 are deleted, and the two bridging benzene rings will be determined from scratch. These bridging benzene rings, as well as the other missing C atoms, will be determined by the so-called connectivity-guided sR1 method, which is introduced in the next paragraph.

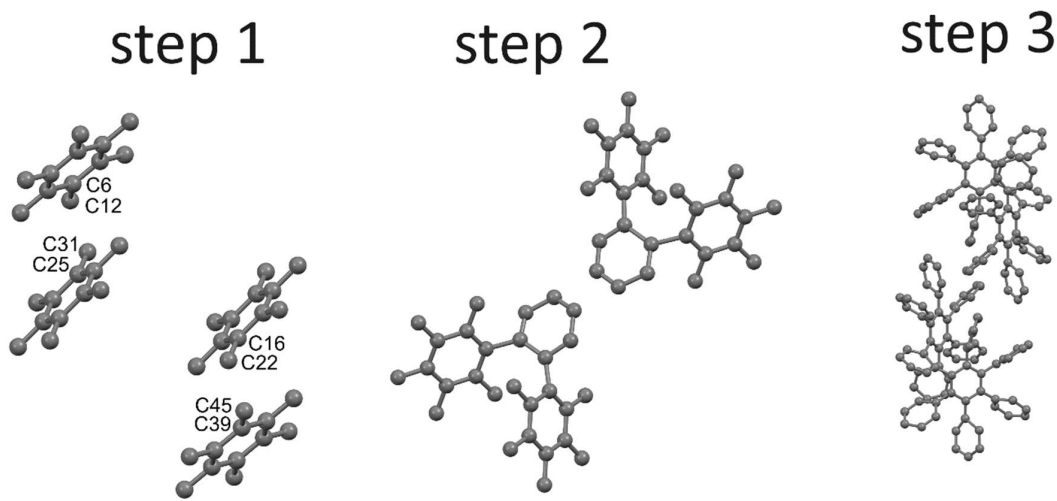

**Figure 1** Assembling the structure of sample 1. Step 1: use the pR1 method to orient and position four benzene-star fragments. Step 2: after deleting C12, C31, C22 and C45, use the connectivity-guided sR1 method to extend six C atoms from C25 and from C39, respectively. Step 3: use the connectivity-guided sR1 method to complete twenty incomplete benzene rings.

The idea of the connectivity-guided sR1 method is simple. Suppose that atom A has been determined and Q is an undetermined atom which is expected to bond to atom A with approximate bond length $r$. A grid has been set up within the unit cell with step size 0.4 Å. To coarsely locate atom Q, only the grid points that are located within a spherical shell, which is centered at atom A and has radii $r - \Delta r \sim r + \Delta r$, need to be tested. Here $\Delta r$ takes the value either 0.3, or 0.4, or 0.5 Å (in this report, 0.5 Å has been used). Other than limiting the initial search range for positioning atom Q, the rest of the connectivity-guided sR1 calculation is same as the normal sR1 procedure (Zhang & Donahue, 2024).

Here is a description on how the connectivity-guided sR1 method is used to determine the two bridging benzene rings. After deleting C12, C31, C22 and C45, one starts from C25 to determine the first bridging benzene ring. The approximate C-C bond length is 1.39 Å. One uses the connectivity-guided sR1 to determine an atom that is bonded to C25, and then uses the same method to determine another atom that is bonded to the newly determined atom, and so on. Doing in sequence, after extending 6 C atoms, a bridging benzene ring is completed. The other bridging benzene ring can be similarly determined by starting from C39. In fact, the determination of both rings can be planned beforehand, and the calculation can be done in one run. Step 2 of figure 1 shows the result after determining these bridging benzene rings.

Step 2 of figure 1 shows that there are 20 single C atoms sticking out from 4 benzene rings. One needs to extend 5 C atoms to each sticking out C atom to complete a benzene ring at that location. The plan to complete these twenty incomplete benzene rings is made beforehand, and in one run the whole structure is completed by the connectivity-guided sR1 calculation. Step 3 of figure 1 shows the completed model of sample 1 after finishing this plan. This model, when compared to the correct model, has all 156 C atoms located within 0.5 Å.

### 3. Assembling the structure of sample 2 by the pre-knowledge guided pR1 and sR1 calculation

Step 3 of figure 2 shows the expected structure of sample 2. This structure also contains one benzene-star fragment. To start assembling this structure, one places a benzene-star fragment that is correctly oriented by the pR1 method such that its center is at (0.3, 0.3, 0.3). Step 1 of figure 2 shows this result.

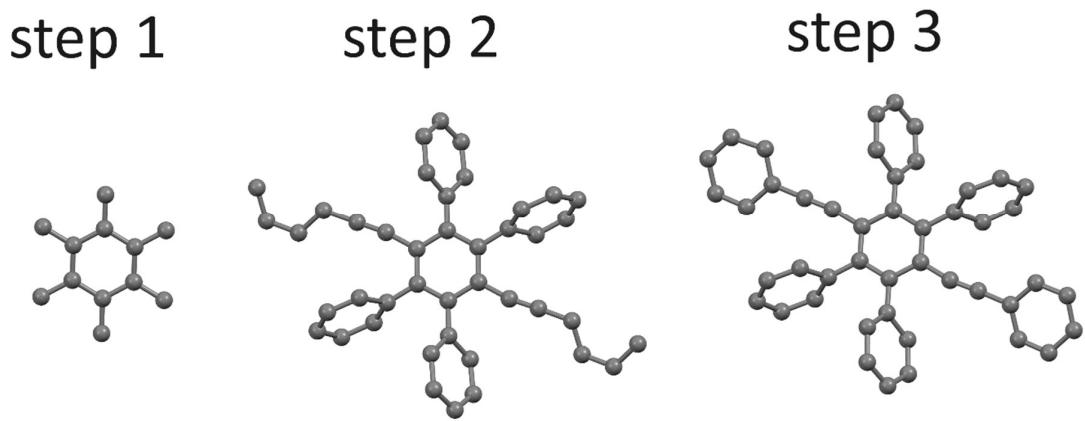

**Figure 2** Assembling the structure of sample 2. Step 1: position a benzene-star fragment that is correctly oriented by the pR1 method such that its center is at (0.3, 0.3, 0.3). Step 2: extend five C atoms from each sticking-out C atom by the connectivity-guided sR1 method. Step 3: complete four additional missing C atoms by the connectivity-guided sR1 method.

The remaining missing C atoms can be determined by the connectivity-guided sR1 method. As seen in step 1 of figure 2, there are six single C atoms sticking out from the central benzene ring. At each sticking out C atom one needs to extend either five or seven C atoms to make a complete benzene ring or to make a benzene ring that is connected to the sticking out atom through an additional C atom, respectively. Because it is unknown where the seven-atom extension should be performed, it is not possible to plan beforehand to complete all the missing atoms in one run. To resolve this ambiguity, one may plan to uniformly extend five atoms from each sticking out C atom. Step 2 of figure 2 shows the result after this planned run. Now it is clear where the additional four missing C atoms are located, and a new plan can be made to locate these missing atoms by a new connectivity-guided sR1 run. Step 3 of figure 2 shows the final completed model after finishing this new plan. This model, when compared to the correct model, has all 46 C atoms located within 0.5 Å.

## 4. Assembling the structure of sample 3 by the normal and the connectivity-guided sR1 methods

The expected structure of sample 3 can be seen in step 3 of figure 3. The structure consists of four pairs of cation and anion. The anion is $I^-$. The cation consists of a heavy atom core (made of three Mo atoms and 13 S atoms) and three $CN^iBu_2$ side chains. The first step of assembling this structure is to use the normal sR1 method to determine the heavy-atom substructure. This calculation has been carried out; however, the result has the following minor problem in atomic type assignment: 4 I are misassigned as Mo, 3 Mo are misassigned as I, one Mo is misassigned as S, one S is misassigned as I, and one N is misassigned as S. These misassignments are corrected and the corrected result is shown in step 1 of figure 3. To this point, all 4 I atoms, all 12 Mo atoms, and 51 S atoms are determined. One S atom is still missing, and one light atom (a nitrogen atom) has already been determined. The remaining missing atoms can be determined by the connectivity-guided sR1 method. The missing S atom is determined by starting from an S atom that the missing atom is expected to bond with and by using an approximate S-S bond length of 2.0 Å. A missing C atom that at one end bonds to two S atoms and at the other end bonds to the N atom that has already been located is determined by starting from one S atom that the missing C atom is expected to bond with and by using approximate S-C bond length 1.8 Å. The remaining two $^iBu$ groups of this side chain are extended by using an approximate bond length of 1.39 Å for N-C or C-C bonds. In a similar way, other $CN^iBu_2$ side chains are determined. Step 2 of figure 3 shows the partial model after locating the missing S atom and completing all three $CN^iBu_2$ side chains of one cation. Determination of the remaining 9 $CN^iBu_2$ side chains of the other three cations can be planned beforehand and be completed in one run of the connectivity-guided sR1 calculation. Step 3 of figure 3 shows the final completed model after finishing this plan. This model, when compared to the correct model, has 183 out of 188 atoms located within 0.5 Å.

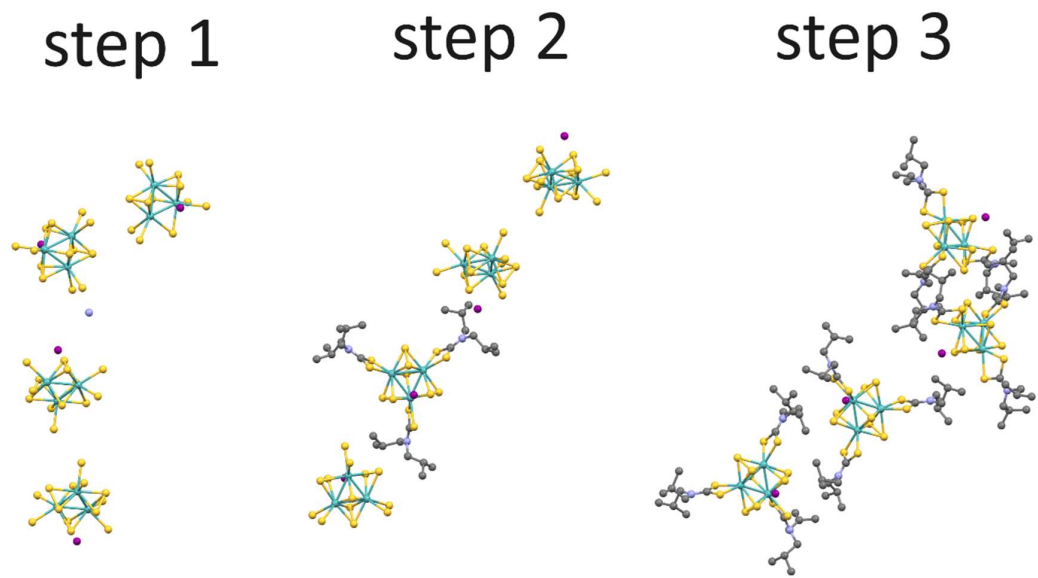

**Figure 3** Assembling the structure of sample 3. Step 1: use the normal sR1 method to determine the heavy-atom substructure. Step 2: use connectivity-guided sR1 to add one missing S atom and to complete three CN$^i$Bu$_2$ side chains of one cation. Step 3: use connectivity-guided sR1 to add 9 missing CN$^i$Bu$_2$ side chains to the other three cations.

## 5. Discussions and conclusions

Sample 1 represents a typical light-atom-only crystal structure. The strategy of first using the pR1 method to orient and position four benzene-star fragments and then completing the model by the connectivity-guided sR1 method is a well-organized way to assemble this structure. However, there is one major drawback in this strategy, namely that the pR1 calculation goes too slow. Using Microsoft Surface Pro 7 and Python programming, it takes the pR1 method 780 seconds to find the orientations and 3700 seconds to position the fragments. In addition, the connectivity-guided sR1 method needs 446 seconds to complete the model (in comparison, 710 seconds if this step is done by the normal sR1 method). In total, the full model is assembled in 4926 seconds. In comparison, starting from scratch, the normal sR1 method can solve the structure in 1060 seconds. Obviously, when there is a choice between the pR1 method and the sR1 method, the pR1 method should be avoided. However, when the data resolution is extremely low, the assistance of the pR1 method becomes necessary (and, fortunately, at a low resolution the number of reflections is also small, and so, the burden of calculation is small, making it much affordable to run the pR1 calculation). For sample 1, when data resolution is truncated to 1.2 Å, the sR1 method alone can no longer solve the structure, but with the pR1 method's help to orient and position four benzene-star fragments, the sR1 method can then complete the model. On the other hand, one should also notice that, at higher resolution, after using normal sR1 method to determine a framework of the structure, the connectivity-guided sR1 is still useful to complete the model, because it goes faster than the normal sR1 method (see above in this paragraph), and more importantly, it can target specific missing atoms (at times these specific missing atoms are hard to get by the normal sR1 method).

Sample 3 represents a typical heavy-atom-containing structure. The strategy of first using the normal sR1 method to determine the heavy-atom substructure and then using the connectivity-guided sR1 method to complete the model is preferred because it goes faster and results in better quality of a model (than a strategy of using the normal sR1 method only). The total calculation time of this strategy to assemble the structure of sample 3 is 1310 seconds, and the resulting model only has 5 atoms being misplaced. In comparison, if the calculation is done by the normal sR1 method all the way through, the total calculation time is 3840 seconds, and the resulting model has 18 atoms being misplaced.

Based on above observations, the main conclusions of this study can be summarized as follows: (1) The usual strategy of assembling a light-atom-only structure is to use the normal sR1 method to construct a framework of the structure and then to use the connectivity-guided sR1 method to complete the model. However, when the data resolution is low, it is necessary to use the pR1 method to assemble known fragments in the first step. (2) The preferred strategy of assembling a heavy-atom-containing structure is to use the normal sR1 method to build the heavy-atom substructure and then to use the connectivity-guided sR1 method to complete the model.

**Acknowledgements**     Professor Robert A. Pascal has kindly provided the data sets of samples 1 and 2.

**References**

Bricogne, G. (1992). *Proceedings of the CCP4 Study Weekend. Molecular Replacement*, edited by W. Wolf, E. J. Dodson & S. Gover, pp. 62-75. Warrington: Daresbury Laboratory.

Burla, M. C., Carrozzini, B., Cascarano, G. L., Giacovazzo, C. & Polidori, G. (2020). *Acta Cryst.* D**76**, 9-18.

Caliandro, R., Carrozzini, B., Cascarano, G. L., Giacovazzo, C., Mazzone, A. & Siliqi, D. (2009). *Acta Cryst.* A**65**, 512-527.

Crowther, R. A. (1972). In *The Molecular Replacement Method*, Ed. M. G. Rossmann, Gorden & Breach, New York, pp. 173-178.

Gorelik, T. E., Lukat, P., Kleeberg, C., Blankenfeldt, W. & Mueller, R. (2023). *Acta Cryst.* A**79**, 504-514.

McCoy, A. J. (2004). *Acta Cryst.* D**60**, 2169-2183.

McCoy, A. J., Grosse-Kunstleve, R. W., Adams, P. D., Winn, M. D., Storoni, L. C. & Read, R. J. (2007). *J. Appl. Cryst*. **40**, 658–674.

McCoy, A.J., Oeffner, R. D., Wrobel, A. G., Ojala, J. R., Tryggvason, K., Lohkamp, B. & Read, R. J. (2017). Ab initio solution of macromolecular crystal structures without direct methods. *Proc. Natl. Acad. Sci. USA* **114**, 3637-3641.

Read, R. J. (2001). *Acta Cryst.* D**57**, 1373-1382.

Read, R. J. & McCoy, A. J. (2016). *Acta Cryst.* D**72**, 375-387.

Rossmann, M. G. (1972). Editor. *The Molecular Replacement Method*. New York: Gordon & Breach.

Rossmann, M. G. (1990). *Acta Cryst.* A**46**, 73-82.

Rossmann, M. G. & Blow, D. M. (1962). *Acta Cryst*. **15**, 24-31.

Rossmann, M. G., Blow, D. M., Harding, M. M., & Coller, E. (1964). *Acta Cryst.* **17**, 338-342.

Zhang, X. & Donahue, J. P. (2024). *Acta Cryst.* A**80**, 237-248.

# Supporting information

### S1. Information on data collection of the samples

All crystals were coated with paratone oil and mounted on the end of a nylon loop attached to the end of the goniometer. Data were collected at 150 K under a dry $N_2$ stream supplied under the control of an Oxford Cryostream 800 attachment. The data collection instrument was a Bruker D8 Quest Photon 3 diffractometer equipped with a Mo fine-focus sealed tube providing radiation at λ = 0.71073 nm or a Bruker D8 Venture diffractometer operating with a Photon 100 CMOS detector and a Cu Incoatec I microfocus source generating X-rays at λ = 1.54178 nm.

### S2. Our implementation of the Wilson method for determination of the parameter B

The raw reflection intensities $F_o^2(hkl)$ are ranked by s=sinθ/λ and are divided into either 20 groups if total number of reflections exceeds 10000 or 10 groups otherwise. For each group, we calculate the average intensity which we denote as $<F_o^2>$, and we also calculate the average s value which we denote as $<s>$. This leads to the following approximate equation:

$$k <F_o^2> \approx [\sum_j f_j^2(<s>)]\exp(-2B<s>^2)$$

In this equation, k is a scaling factor for the observed intensities, and B is the parameter whose value we are seeking for.

This equation can convert into a linear form of Y=a+bX, in which X and Y are defined as:

$$X = <s>^2$$

$$Y = ln\frac{<F_o^2>}{\sum_j f_j^2(<s>)}$$

and the intercept a and the slope b are:

$$a = -\ln(k)$$

$$b = -2B$$

After calculating the (X,Y) data points for all the groups we use linear fit to find the intercept a and the slope b, and further calculate the parameters k and B as follows:

$$k = \exp(-a)$$

$$B = -b/2$$

Once the parameter B is obtained, a sharpened intensity is calculated as:

$$(F_o^2)_{sharpened} = F_o^2 \exp(2Bs^2)$$

However, the sharpened intensities are not scaled by multiplying with the scaling factor k; instead, they are scaled such that their sum equals $\Sigma_{hkl} \Sigma_j f_j^2(hkl)$.

### S3. The algorithm of comparing two models

This algorithm only compares the positions of the atoms in the two models; it disregards the types of atoms. Let a, b, and c be the length of the unit cell edges. One difficulty of making this comparison is that the two models may have different locations relative to the unit cell edges. The algorithm uses an approximate method to overcome this difficulty: shift all atoms of one model (model A) such that one of its atoms overlaps one atom of the other model (model B). Try this for all atoms in both models. So, the calculation needs to be repeated for $N_A \times N_B$ times, where $N_A$ and $N_B$ are the number of atoms in the models A and B, respectively. Further these calculations also need to be repeated after inverting model B, in case model A matches the inverted version of B better. After repeating the calculation, use the result of the best match that has been discovered.

After overlapping one particular atom of model A with a particular atom of model B by properly shifting all atoms of model A, we make sure all atoms of both models are located within the unit cell (by adding 1 to or subtracting 1 from the fractional coordinates x,y,z of an atom). With such preparation we are ready to count how many atoms of A are each within s = 0.5 Å of an atom in B. To make such comparison quickly, we convert the fractional coordinates x,y,z of an atom to integers in following way:

$N_x = int(a/s)$, $N_y = int(b/s)$, $N_z = int(c/s)$

$I_x = int(xN_x)$, $I_y = int(yN_y)$, $I_z = int(zN_z)$

Thus, $(I_x, I_y, I_z)$ is the integer coordinate of an atom.

Prepare a mask $M_A(i, j, k)$ for model A, where i=0 to $N_x$-1, j=0 to $N_y$-1, and k=0 to $N_z$-1. $M_A(i, j, k)$ takes value 0 unless (i,j,k) is the integer coordinate of an atom. Similarly, $M_B(i, j, k)$ is a mask for model B.

The number of atoms of model A which overlaps within 0.5 Å with an atom of B can be calculated as $\sum_{i,j,k} M_A(i,j,k) M_B(i,j,k)$.

### S4. Technical details of using pR1 to determine the orientation of a fragment as well as to position a fragment of a known orientation

A local Cartesian coordinate system is set up for the fragment by selecting a local origin and three orthogonal directions.

A Cartesian coordinate system is also set up for the unit cell. The origin of this system overlaps the origin of the unit cell. Its three unit vectors **x**, **y**, and **z** are related to the cell vectors **a** and **b** as follows: **x** = **a**/|**a**|, **y** = (**b**-**b**·**xx**)/| **b**-**b**·**xx** |, **z** = **x**×**y**. The Cartesian coordinates and the fractional

coordinates of an atom in the unit cell are inter-converted during the calculations. Using cell parameters $a$, $b$, $c$, $\alpha$, $\beta$, and $\gamma$, the relation between fractional coordinates $(x_f, y_f, z_f)$ and Cartesian coordinates $(x_c, y_c, z_c)$ is expressed as:

$$\begin{pmatrix} x_c \\ y_c \\ z_c \end{pmatrix} = \begin{pmatrix} a & b\cos\gamma & c\cos\beta \\ 0 & b\sin\gamma & \dfrac{c(\cos\alpha - \cos\beta\cos\gamma)}{\sin\gamma} \\ 0 & 0 & \dfrac{V}{ab\sin\gamma} \end{pmatrix} \begin{pmatrix} x_f \\ y_f \\ z_f \end{pmatrix}$$

in which the cell volume V is calculated as:

$$V = abc\sqrt{1 - \cos^2\alpha - \cos^2\beta - \cos^2\gamma + 2\cos\alpha \times \cos\beta \times \cos\gamma}$$

At start, a fragment is positioned such that its local Cartesian system overlaps the Cartesian system of the unit cell. The fragment is attached to its local Cartesian system. So, rotation and translation of the fragment is realized by rotating and translating its local Cartesian system in the cell Cartesian system. Translation is realized by translating the local origin in the cell Cartesian system (or rather in the cell fractional coordinate system, and the fractional coordinates and the cell Cartesian coordinates are inter-converted). A general rotation is consisted by three simple rotations: a rotation around x-axis (of the local frame) in the direction from y-axis to z-axis through angle $\psi$; a rotation around z-axis (of the local frame) in the direction from x-axis to y-axis through angle $\varphi$; and another rotation around x-axis (of the local frame) in the direction from y-axis to z-axis through angle $\zeta$. Before rotation, a point has Cartesian coordinates $(x,y,z)$. After a rotation of angles $(\psi,\varphi,\zeta)$, the point moves to a new location in the same Cartesian system with Cartesian coordinates $(x',y',z')$, which are calculated by:

$$\begin{pmatrix} x' \\ y' \\ z' \end{pmatrix} = \begin{pmatrix} 1 & 0 & 0 \\ 0 & \cos\psi & -\sin\psi \\ 0 & \sin\psi & \cos\psi \end{pmatrix} \begin{pmatrix} \cos\varphi & -\sin\varphi & 0 \\ \sin\varphi & \cos\varphi & 0 \\ 0 & 0 & 1 \end{pmatrix} \begin{pmatrix} 1 & 0 & 0 \\ 0 & \cos\zeta & -\sin\zeta \\ 0 & \sin\zeta & \cos\zeta \end{pmatrix} \begin{pmatrix} x \\ y \\ z \end{pmatrix}$$

In general, to cover all possible rotations, the range of $\psi$ should be 0 to 360 degrees, the range of $\varphi$ should be 0 to 180 degrees, and the range of $\zeta$ should be 0 to 360 degrees. When a fragment has n-fold rotation symmetry and the rotation axis is along the x-axis of its local Cartesian system, then the range of $\zeta$ can reduce to 0 to 360/n degrees.

The space spanned by rotation and translation is a 6-dimensional orientation-location space. (It is 5-dimensional if the fragment is linear.) To coarsely locate the global minimum point or the local minimum points (the holes) of a pR1 map in this 6-dimensional space, a grid is set up with 0.4 Å step size in translations within the cell and 5-degree step size in rotation angles. The range of the rotation angles are: 0 to 360 degrees for $\psi$, 0 to 180 degrees for $\varphi$, and 0 to 360 degrees for $\zeta$. If the fragment has n-fold rotation symmetry, and the rotation axis is arranged along the local x-axis, the range of $\zeta$ can reduce to 0 to 360/n. The precision of locating the global minimum point or the local minimum points is refined by halving the step size locally five times.

The general hypothesis is that the deepest hole of a pR1 map in a 6-dimensional orientation-location space determines the orientation and location of a missing fragment (Zhang & Donahue, 2024). Note that by "location of a fragment" we mean the location of the local origin of the fragment (while keeping its orientation unchanged); similarly, "locating a fragment" means locating the local origin of a fragment. Determining pR1 holes in a 6-dimensional space is time-consuming. To cut calculation time, the problem is divided into two 3-dimensional calculations, that is, finding the orientation and the location in two separate steps.

For a free-standing fragment, that is, a model consisting of a single fragment with no other known atoms, the pR1 only depends on the orientation of the fragment. Therefore, the possible orientations of all missing fragments are detected by the holes in this pR1 map of a free-standing fragment in a 3-dimensional orientation space. Due to the possible symmetry of a fragment and the redundancy in representing an orientation with the three rotation angles, many orientation representations are equivalent to each other. Considering two orientation representations, if the atoms of the fragment of one orientation representation can match the atoms of the fragment of the other representation in a one-to-one basis within 0.25 Å (see the algorithm for this type of comparison in section S3), the two representations are considered equivalent. The non-equivalent orientation representations are filtered out from all detected representations. These serve as the candidate orientations.

The candidate orientations are ranked from the smallest R1 to the highest R1 and are labelled by 0, 1, 2, etc. In some situations, for example, if it is known that there are only two fragments in the structure, then it is obvious that one fragment has orientation 0 and the other has orientation 1. In such a situation, fragment 0 takes orientation 0, and its location is determined by the deepest hole in a pR1 map of a 3-dimensional location space. After determination of fragment 0, a new pR1 is defined by including the atoms of fragment 0 as known atoms and taking orientation 1 the location of fragment 1 is determined by the deepest hole of this new pR1 map in a 3-dimensional location space. In other situations, there are multiple fragments in the structure. In such situations, to determine a missing fragment, it is necessary to try all possible orientations. For each trial orientation, the best choice of location of a missing fragment is determined by the deepest hole in a pR1 map of a 3-dimensional location space. This best choice is combined with the trial orientation to form a candidate orientation-location. Trying all candidate orientations leads to a list of candidates of orientation-locations which can be ranked from the smallest R1 to the highest R1. The missing fragment is determined by the candidate orientation-location of the lowest R1. With this missing fragment being determined, a newer pR1 is defined by including the atoms of the newly determined fragment as known atoms. This newer pR1 is used to determine the next missing fragment in the same way.

When using the single atom R1 (sR1) to locate single missing atoms (Zhang & Donahue, 2024), a rule excluding clustering ghost atoms and a rule excluding triangular bonding are enforced. Similarly, here, when using pR1 to locate a missing fragment, these rules are also enforced: if a trial orientation-location for a missing fragment will cause one of its atoms being a clustering ghost atom, or one of its atoms will involve triangular bonding with two known atoms, this trial is disqualified as a candidate for determining the orientation-location of the missing fragment.

**References**

Zhang, X. & Donahue, J. P. (2024). *Acta Cryst.* A80, 237-248.